\documentclass[journal,letterpaper,twocolumn,final,10pt]{IEEEtran}
\usepackage{amsthm}
\usepackage{amsmath}
\usepackage{amsfonts}
\usepackage{dsfont}
\usepackage{mathrsfs}
\usepackage{pifont}
\usepackage{amssymb}
\usepackage{verbatim}
\usepackage{upgreek}
\usepackage{color}
\usepackage[ruled]{algorithm2e} 
\usepackage[final]{graphicx}
\usepackage{epsfig}
\usepackage{eucal}
\usepackage{cite}
\usepackage{subfigure}
\usepackage{cuted}
\usepackage{caption}
\usepackage{multirow}

\newcommand{\bbm}{\begin{bmatrix}}
\newcommand{\ebm}{\end{bmatrix}}

\newcommand{\bit}{\begin{itemize}}
\newcommand{\eit}{\end{itemize}}

\newcommand{\ben}{\begin{enumerate}}
\newcommand{\een}{\end{enumerate}}

\newcommand{\bdesc}{\begin{description}}
\newcommand{\edesc}{\end{description}}

\newcommand{\bea}{\begin{array}}
\newcommand{\eea}{\end{array}}

\newcommand{\beqa}{\begin{eqnarray}}
\newcommand{\eeqa}{\end{eqnarray}}

\newcommand{\ds}{\displaystyle}

\newcommand{\Comment}[1]{}

\def\N{{\mathds N}}

\def\R{{\mathds R}}

\def\cA{\mbox{$\mathcal A$}}
\def\cB{\mbox{$\mathcal B$}}

\def\cL{\mbox{$\mathcal L$}}
\def\cN{\mbox{$\CMcal N$}}

\newcommand{\be}{\begin{equation}}
\newcommand{\ee}{\end{equation}}

\newcommand{\btheta}{{\mbox{\boldmath $\theta$}}}

\newcommand{\dmin}{\begin{displaystyle}\min\end{displaystyle}}

\title{EM-based Algorithm for Unsupervised Clustering of Measurements from a Radar Sensor Network}

\author{Linjie Yan, Pia Addabbo, \IEEEmembership{Senior Member, IEEE}, Nicomino Fiscante, 
Carmine Clemente, \IEEEmembership{Senior Member, IEEE}, Chengpeng Hao, \IEEEmembership{Senior Member, IEEE}, 
Gaetano Giunta, \IEEEmembership{Senior Member, IEEE}, and Danilo Orlando, \IEEEmembership{Senior Member, IEEE}
\thanks{This work was supported in part by the National Natural Science Foundation of China under Grants no.
62201564 and 61971412.}
\thanks{Linjie Yan and Chengpeng Hao are with the Institute of Acoustics, Chinese Academy 
of Sciences, Beijing, 100190, China. E-mail: {\tt yanlinjie16@163.com; haochengp@mail.ioa.ac.cn}.}
\thanks{Pia Addabbo is with Universit\`a degli Studi ``Giustino Fortunato", viale Raffale Delcogliano, 12, 82100 Benevento, Italy. E-mail: {\tt p.addabbo@unifortunato.eu} }
\thanks{Nicomino Fiscante and Gaetano Giunta are with Industrial, Electronic and Mechanical Engineering Department, University of Roma Tre, 00146 Rome, Italy. 
E-mail: {\tt nicomino.fiscante@virgilio.it; gaetano.giunta@uniroma3.it}.}
\thanks{Carmine Clemente is with the University of Strathclyde, Department of Electronic and Electrical Engineering, 204 George Street, 
G1 1XW, Glasgow, Scotland. E-mail: {\tt carmine.clemente@strath.ac.uk}}
\thanks{Danilo Orlando is with the Engineering Department of Universit\`a degli Studi ``Niccol\`o Cusano'', via Don Carlo Gnocchi 3, 
00166 Roma, Italy. E-mail: {\tt danilo.orlando@unicusano.it}.}
}

\begin{document}

\maketitle

\begin{abstract}
This paper deals with the problem of clustering data returned by a radar sensor network that
monitors a region where multiple moving targets are present. The network is formed by nodes
with limited functionalities that transmit the estimates of target positions (after a detection) 
to a fusion center without any association between measurements and targets. To solve the problem at hand,
we resort to model-based learning algorithms and instead of applying the plain maximum likelihood 
approach, due to the related computational requirements,
we exploit the latent variable model coupled with the expectation-maximization algorithm.
The devised estimation procedure returns posterior probabilities that are used
to cluster the huge amount of data collected by the fusion center. Remarkably, we also
consider challenging scenarios with an unknown number of targets and estimate it by means of
the model order selection rules. The clustering performance of the proposed strategy
is compared to that of conventional data-driven methods over synthetic data. The numerical
examples point out that the herein proposed solutions can provide reliable clustering
performance overcoming the considered competitors.
\end{abstract}


\section{Introduction}
\label{Sec:Introduction}
\IEEEPARstart{I}n the recent years, the increase of computational resources 
has heavily promoted the development and the implementation of sophisticated 
signal processing techniques in real radar systems.
More importantly, with the advent of the big data era, statistical signal processing 
algorithms have been incorporated into a wider class called ``Machine Learning'' that has become 
more and more popular over the years \cite{murphy2012machine,theodoridis2015machine} and also 
comprises deep learning techniques.
Two main design approaches can be identified within this wide class: data-driven and model-based oriented designs. 
The former operates through a learning stage  that relies on training data, namely a set of 
input-output pairs used to learn the algorithm from the available data, whereas 
the latter relies on a model grounded on the first physics principles.
In both cases, the exploitation of high performance boards becomes unavoidable 
to fulfill the tight (time) requirements of real radar systems.
The methods devised in this paper are framed in the model-based class.

In parallel with the technological advancements, the operating scenarios 
have become more and more challenging as well as the corresponding estimation/optimization problems.
A related example is represented by scenarios where swarms 
of (possibly noncooperative) targets are moving in the region of interest 
that is monitored by a network of passive/active radars 
\cite{Chernyak,multistaticSonar,10154135,8168427,8681707,6905829,6644279,BOUTKHIL2018455,https://doi.org/10.1049/rsn2.12358,5089551}. 
As a matter of fact, the advantages arising from the use of 
a sensor network are well-known and we mention here the spatial diversity and the energy integration.
For this reason, in modern radar systems, the cooperation between 
systems distributed over the region under surveillance becomes a key factor 
to improve the reliability of the entire surveillance system. For instance, with 
focus on ground radars, the spatial diversity of the sensor network can be exploited 
to face deception/saturation jamming techniques such as range gate stealing \cite{EW101,stimson2014stimson}, 
that generate false targets to make the radar lose range track on the target.

However, radar networks require special attention in handling a huge amount of data 
provided by each node, especially in the presence of multiple moving targets 
\cite{JAVADI202048,7585817,10.1049/ip-f-1.1983.0078}. In this case, the 
detection can be either centralized or decentralized.
In centralized detection, sending raw observations from radar sensors to the  
fusion center (FC), where the final decision process takes place, imposes 
a large communication burden. On the other hand, in decentralized detection, radar
sensors take their local decisions on the presence of prospective targets
and, then, transmit the results of such decisions (compressed data) to the FC. In this case,
the amount of transmitted data is lower than in the previous configuration with a consequent
energy saving but at the price of performance loss. As a matter of fact,
the overall system does not take advantage of diversity.
The design of optimized decentralized detector network, which entails
the design of both optimal local detectors and fusion rules at FC, 
does not  represent an easy task \cite{Tenney1981501} and some strategies to solve 
this difficult problem have been proposed in the past \cite{Fang20095682,NIU2006380,https://doi.org/10.1049/iet-wss.2013.0116}.
Moreover, in radar networks, the tracking of multiple targets represents a very challenging 
problem due to the limited data computational capacity of the FC as well as of the transmit 
energy of each radar \cite{https://doi.org/10.1049/iet-wss.2013.0116,YAN2020173,5338565,bar1995multitarget}. 
Generally speaking, this implies the necessity of using sophisticated (centralized or decentralized) 
fusion methods \cite{nannuru2016multisensor,yang2015multi,chang1997optimal,battistelli2013consensus}. 
Anyway, the implementation of centralized fusion methods is computationally 
expensive \cite{nannuru2016multisensor}. Nevertheless, a preliminary stage
that preprocesses and clusters the received observations into homogeneous measurement sets,
might reduce the computational load at the FC \cite{li2018robust}

Clustering algorithms have been also applied for the design 
of CFAR and/or selective/robust detectors \cite{9112312,9926070}. 
Particularly, in \cite{9112312}, received data are transformed to generate
specific features based upon the maximal invariant statistic \cite{scharf1991statistical} for that problem.
Then, such features are clustered in a two-dimensional plane to come up
with a detector that is invariant to the disturbance parameters and, hence,
can guarantee the constant false alarm rate property.
Another design methodology, based on the previous approach, 
is proposed in \cite{9926070}, where sub-optimal strategies with low complexity
have been developed.
The expectation-maximization (EM) algorithm is also used for clustering data that 
can be modeled as a Gaussian Mixture\cite{murphy2012machine}. In particular, in the context of 
radar systems, the EM algorithm is used to partition clutter returns based upon
their spectral properties \cite{9321174}. To this end, fictitious hidden random variables are introduced
to represent the clutter type of each range bin. Interestingly, 
the classification algorithms have been devised by accounting for different models of the clutter
covariance matrix.
Whereas, in \cite{10058041}, joint detection and classification architectures 
have been proposed by extending the work of \cite{9321174} to the case
where an unknown number of multiple point-like targets are present in the region of interest.
In \cite{10274860}, the EM algorithm is still used for the target detection
in heterogeneous clutter scenarios.
 
In this paper, we provide a solution to the problem of handling multiple targets in a scene. Actually, this issue 
is getting more and more critical with the growing number of scenarios in which multiple targets need
to be monitored in crowded spaces.  As a matter of fact, conventional radar scenarios 
are getting more densely populated with the arrival of new classes of  targets  
such as unmanned/autonomous vehicles (either in air, land, or sea). The same remark also holds for
less conventional but equally challenging radar scenarios such as monitoring space targets.
Therefore, we focus on a radar sensor network whose monostatic nodes 
illuminate the same region of interest where an unknown number of targets are moving.
Each node does not perform any association between
the measurements related to a detection and the detected targets, and 
sends to the FC the position estimates corresponding to each detection within a common observation time window.
At the design stage, we assume that measurements are collected over a time interval such that
the target trajectories can be approximated as straight lines (with a not necessarily constant
velocity). Assuming a specific distribution for the measurement noise, we exploit the likelihood
function of data to solve the clustering problem. In this respect, we do not resort to the
maximum likelihood principle, because it requires the evaluation of
the joint likelihood function for each partition of the entire
measurement set and targets' number. It is clear that such a task is unacceptable from a computational 
standpoint and, more importantly, the maximum likelihood approach
would return estimates corresponding to the maximum allowed model order since
the likelihood function
monotonically increases with the number of unknown parameters \cite{kay2005multifamily,Stoica1}. 
Therefore, to circumvent the above limitations, we introduce fictitious and 
unobserved discrete random variables
that represent target labels associated with the measurements gathered by the FC.
Then, exploiting the joint
distribution of measurements and the hidden labels, we develop an estimation procedure
grounded on the EM algorithm \cite{Dempster77} that allows us
to obtain a nondecreasing sequence of likelihood values as well as closed-form expressions
for the updates of the estimates.
The clusters are formed by applying the maximum a posteriori rule, 
whereas an estimate of the number of targets is returned through the Model Order Selection (MOS) rules  
\cite{StoicaBabu,StoicaBabu1,Stoica1,Kay2005,kay2005multifamily}.

The performance analysis is carried out over synthetic data and starts
from the case where the number of targets is known to proceed
with the case where the latter is unknown. As terms of comparison, 
we consider two conventional data-driven algorithms. The numerical examples
show the superiority of the proposed approach over the considered competitors
in classifying the collected measurements.

The remainder of this paper is organized as follows. In the next section,
we describe the surveillance system and provide a formal statement of the problem.
In Section \ref{sec_ParEst_known_L}, we devise the EM-based estimation procedure
when the number of targets is known, whereas in Section \ref{sec_Unknown_L} we extend 
such a procedure to the case of an unknown number of targets. The numerical examples
are contained in Section \ref{Sec:Analysis} and, finally, concluding remarks along with
the description of possible future research lines are confined to Section \ref{Sec:conclusions}.

\subsection{Notation}
In the sequel, vectors are denoted by boldface lower-case.
%The $i$th entry of a vector $\boa$ is represented by $\boa(i)$, whereas
%The $(i,j)$th entry of a matrix $\bA$ is indicated by $\bA(i,j)$.
%Symbols $\det(\cdot)$, $\tr(\cdot)$, $(\cdot)^T$, and $(\cdot)^\dag$ denote the determinant, trace, transpose, 
%and conjugate transpose, respectively.
%Symbol $\|\cdot\|$ denotes the Euclidean norm of a vector. 
As to numerical sets, $\N$ is the set of natural numbers, 
$\R$ is the set of real numbers, and $\R^{N\times M}$ is the Euclidean space of $(N\times M)$-dimensional 
real matrices (or vectors if $M=1$). The Cartesian product of two sets $A$ and $B$ is
denoted by $A\times B$.
%$\R_+^{N\times M}$ is the set of $(N\times M)$-dimensional 
%real matrices (or vectors if $M=1$) whose entries are greater than or equal to zero, 
%$\C$ is the set of 
%complex numbers, and $\C^{N\times M}$ is the Euclidean space of $(N\times M)$-dimensional 
%complex matrices (or vectors if $M=1$). 
%The modulus of a real number $x$ is denoted by $|x|$.
%$\bI$ and $\bzero$ stand for the identity matrix and the null vector or matrix of proper size. 
%Given $a_1, \ldots, a_N \in\C$, $\diag(a_1, \ldots, a_N)\in\C^{N\times N}$ indicates 
%the diagonal matrix whose $i$th diagonal element is $a_i$.
The acronyms PDF and PMF stand for Probability Density Function and Probability Mass Function, respectively, whereas
the conditional pdf of a random variable $x$ given 
another random variable $y$ is denoted by $f(x|y)$. The probability of an event $\cA$ is defined as $P\{\cA\}$, whereas
the conditional probability of $\cA$ given another event $\cB$ is $P\{\cA|\cB\}$.
Symbol $\lfloor \cdot \rfloor$ represents the highest integer lower than the argument while
$|x|$ is the absolute value of $x\in\R$. 
Finally, we write $x\sim\cN(m, \sigma^2)$ if $x$ is a Gaussian random variable 
with mean $m$ and variance $\sigma^2>0$.

\section{Signal Model and Problem Formulation}
\label{Sec:SignalModel}
\begin{figure}[tbp]
\begin{center}
\includegraphics[width=0.85\linewidth]{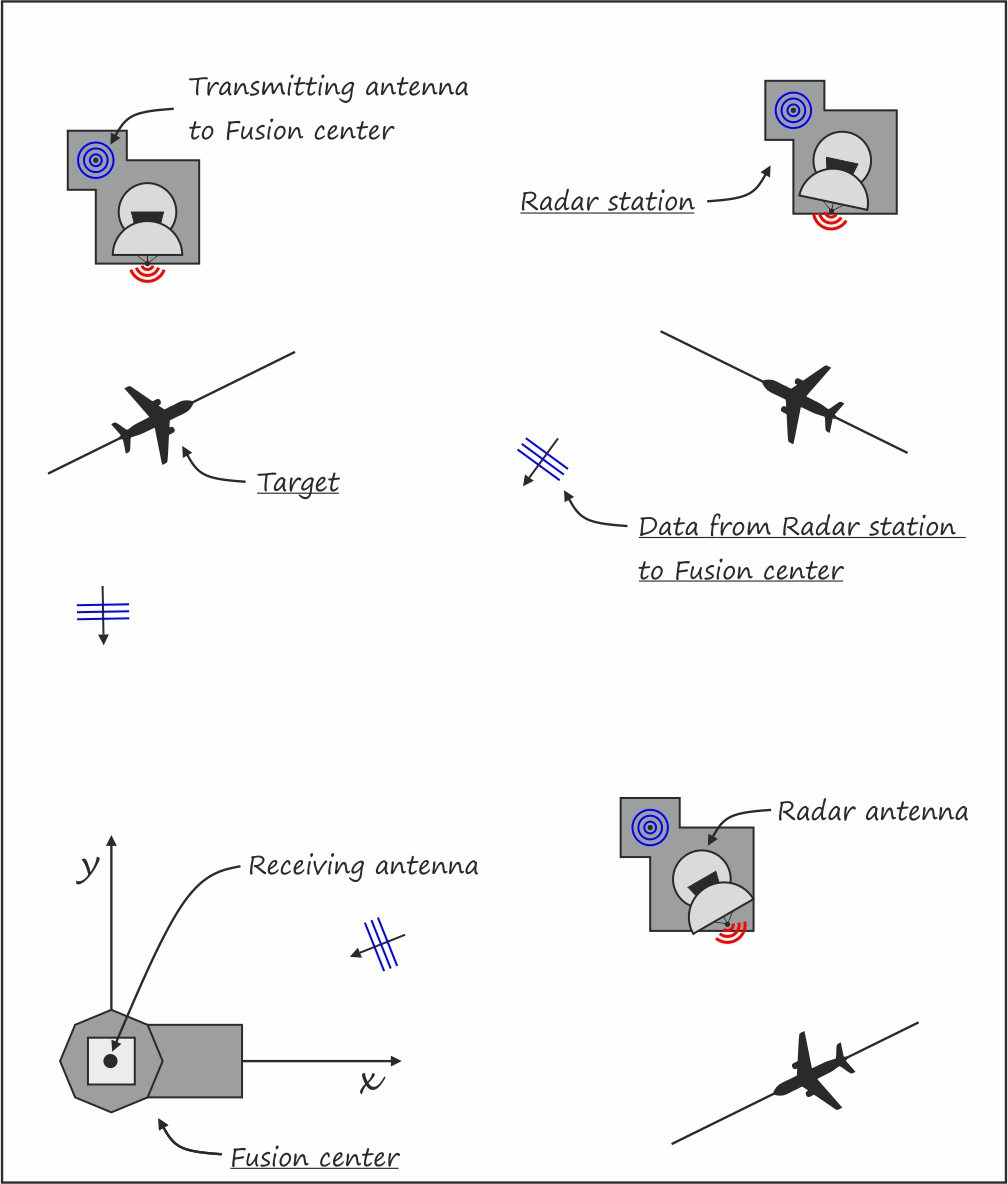}
\caption{System geometry and coordinates.}
\label{systemgeometry}
\end{center}
\end{figure}

Let us consider a sensor network of $K\in\N$ monostatic radars deployed to illuminate the same region of interest.
The capabilities of such systems are limited to target detection and rough estimation of its
range and azimuth. Assuming that the region of interest is populated by an unknown number, $L$ say, 
of multiple point-like moving targets, the measurements obtained by each sensor are transmitted to a fusion center
that converts the received polar coordinates into Cartesian coordinates.
Notice that each radar system works in an asynchronous way with respect to the other systems.
At a given time instant and for the $l$th target, the fusion center receives $0\leq K_l\leq K$
measurements from the nodes that have detected the same target and we denote
the corresponding coordinates by $(x_{l,k},y_{l,k})\in\R\times \R$, $l=1,\dots,L$ and
$k\in\Omega_l\subseteq \{1,\ldots,K\}$, 
where $\Omega_l$ is the set of systems that have detected the $l$th target and measured the related parameters.
The fusion center collects measurements for a preassigned time interval without knowing 
which target is associated with the received measurements.
The first operation performed by the fusion center consists in clustering
data under the assumption that targets' motion over short time intervals can be approximated 
as a straight line. The system geometry and the considered scenario are depicted in Figure \ref{systemgeometry}.
Thus, in the absence of disturbance, we consider the following model for the measurements
\be
y_{l,k}=a_l x_{l,k} + b_l,
\label{eqn:problem1}
\ee
where $a_l\in\R$ and $b_l\in\R$ are unknown coefficients associated with the motion of the $l$th target
over the considered time interval.
It is important to observe that, in principle, the above clustering problem could be solved by using
the maximum likelihood approach. However, the solution would experience a combinatorial computational 
complexity due to the fact that the likelihood function is computed over all the 
possible associations and number of targets.
In addition, the maximum likelihood approach would always select the maximum number of targets since
the likelihood function is monotonically increasing with the number of parameters \cite{Stoica1}.
To overcome the above drawbacks, we resort to the following approaches:
\begin{itemize}
\item the Latent Variable Model \cite{murphy2012machine,9321174,10058041,9799753,10274860} that allows us to reduce the computational 
complexity related to each possible association;
\item the MOS rules \cite{StoicaBabu,StoicaBabu1,Stoica1,Kay2005,kay2005multifamily} 
that mitigate the overfitting inclination of the maximum likelihood approach
for the estimation of the target number.
\end{itemize}

Starting from the first item, let us assume for the moment that the number of targets is known
and denote by $N$ the entire number of measurements, $y_n$ say, collected by the fusion center.
Then, we reformulate the problem at hand by introducing $N$ independent and identically distributed 
hidden discrete random variables, $c_n$ say, taking on values in $\mathcal{A} = \{1,\dots,L\}$, such that 
for the generic $n$th measurement we have that
\be
y_{n}|c_{n}=l \sim \cN(a_l x_{n} + b_l,\sigma^2_l), \quad n=1,\dots,N,
\label{eqn:problem2}
\ee
where the measurement noise related to the $l$th target is modeled as a Gaussian random 
variable with unknown variance $\sigma^2_l>0$.
Notice that random variables $c_n$ represent target identifiers associated with each measurement and, as described below,
can be used to suitably cluster measurements.
Assuming that the measurements are statistically independent and that the PMF of $c_n$ is denoted
by $\pi_{l}=P\{c_{n}=l\}$, $l\in\mathcal{A}$, the joint PDF of $y_{1},\dots,y_{N}$ can be written as:
\begin{equation}
\begin{split}
&f(y_{1},\dots,y_{N};\boldsymbol{\theta},\pi_{1},\dots,\pi_{L})\\
&=\prod_{n=1}^{N} f(y_{n};\boldsymbol{\theta},\pi_{1},\dots,\pi_{L})\\
&=\prod_{n=1}^{N} \sum_{l \in \mathcal{A}} f(y_{n},c_{n}=l;\boldsymbol{\theta},\pi_{1},\dots,\pi_{L})\\
&=\prod_{n=1}^{N} \sum_{l \in \mathcal{A}} f(y_{n}|c_{n}=l;\boldsymbol{\theta})\pi_{l},
\label{eqn:joint_pdf}
\end{split}
\end{equation}
where $\boldsymbol{\theta}=[a_{1},\dots,a_{L},b_{1},\dots,b_{L},\sigma^2_{1},\dots,\sigma^2_{L}]^{T}\in \R^{3L \times 1}$.

Finally, we exploit the above PDF to obtain the estimates of the unknown parameters that will be used to
build up a maximum a posteriori rule allowing for measurement association and, hence, target clustering.
Specifically, the unknown parameters are estimated by solving
\be
\max_{{\bf \theta} \atop \pi_1,\ldots,\pi_L}
\prod_{n=1}^{N} \sum_{l \in \mathcal{A}} f(y_{n}|c_{n}=l;\boldsymbol{\theta})\pi_{l}
\ee
or equivalently
\be
\max_{{\bf \theta} \atop \pi_1,\ldots,\pi_L}
\sum_{n=1}^{N} \log \left[ \sum_{l \in \mathcal{A}} f(y_{n}|c_{n}=l;\boldsymbol{\theta})\pi_l\right].
\label{eqn:estimationProblem}
\ee
Denoting by $\widehat{\btheta}$ and $\widehat{\pi}_l$, $l\in\cA$, the estimates of
$\btheta$ and $\pi_l$, $l\in\cA$, respectively, we associate measurement $y_n$ with
the target identifier $\widehat{l}$ exhibiting the highest a posteriori probability, namely
\be
\widehat{l}=\arg\max_{l\in\cA}
P\left\{c_{n}=l|y_n;\widehat{\btheta},\widehat{\pi}_{l}\right\}.
\label{eqn:MAPclustering}
\ee
In the next section, we design an estimation procedure to solve problem
\eqref{eqn:estimationProblem} and, as a byproduct, \eqref{eqn:MAPclustering}, 
grounded on the EM-algorithm. Then, in Section \ref{sec_Unknown_L}, we address
the case where $L$ is unknown.

\section{Estimation Procedure for Known Number of Targets}
\label{sec_ParEst_known_L}
From a mathematical point of view, the plain maximization in \eqref{eqn:estimationProblem} is a difficult
task at least to the best of authors' knowledge. For this reason, we resort to the EM-algorithm
that is an iterative procedure with closed-form updates for the estimates of interest
and provides at least a local maximum \cite{Dempster77,bishop2007pattern,murphy2012machine}.
The EM-algorithm repeats two steps called E-step and M-step until a stopping
criterion is not satisfied. The former consists
in updating the a posteriori probability of the event $c_n=l$ given the $n$th measurement $y_n$
whereas in the latter step, the log-likelihood function is maximized to obtain
updated parameter estimates.

Thus, let us start form the E-step and denote 
by $\widehat{\boldsymbol{\theta}}^{(h-1)}$ and $\widehat{\pi}_{l}^{(h-1)}$, $l\in\cA$,
the estimates of $\btheta$ and $\pi_l$, $l\in\cA$, at the $(h-1)$th iteration, respectively.
The E-step leads to the computation of 
\begin{align}
p_{n}^{(h-1)}(l)&=P\left\{c_{n}=l|y_n;\widehat{\boldsymbol{\theta}}^{(h-1)},\widehat{\pi}_{l}^{(h-1)}\right\} 
\nonumber
\\ 
&=\frac{f\left(y_{n}|c_{n}=l;\widehat{\boldsymbol{\theta}}^{(h-1)}\right)\widehat{\pi}_l^{(h-1)}}
{\sum\limits_{\bar{l} \in \mathcal{A}} f\left(y_{n}|c_{n}=\bar{l};\widehat{\boldsymbol{\theta}}^{(h-1)}\right)\widehat{\pi}_{\bar{l}}^{(h-1)}}
\label{eqn:E-step}
\end{align}
for $l\in\cA$ and $n=1,\ldots,N$.
As for the M-step, after applying the Jensen inequality to the argument of \eqref{eqn:estimationProblem},
we come up with the following optimization problem
\begin{equation} 
\max\limits_{\boldsymbol{\theta} \atop 
\pi_l, l\in\mathcal{A}}  
\sum_{n=1}^{N} \sum_{l \in \mathcal{A}}p_{n}^{(h-1)}(l)\log\left(\frac{f(y_{n}|c_{n}=l;\boldsymbol{\theta})\pi_l}{p_{n}^{(h-1)}(l)}\right),
\label{eqn:M-step}
\end{equation}
where
\begin{equation} 
f(y_{n}|c_{n}=l;\boldsymbol{\theta})
=\frac{\exp{\left[-\frac{1}{2\sigma^2_l}\left(y_n-a_lx_n-b_l \right)^2\right]}}
{\sqrt{2\pi}\sigma_l}.
\label{eqn:pfd}
\end{equation}
Problem \eqref{eqn:M-step} is tantamount to
\begin{multline} 
\max\limits_{\boldsymbol{\theta} \atop 
\pi_l,l\in \mathcal{A}}  \sum_{n=1}^{N} \sum_{l \in \mathcal{A}} \Big[  p_{n}^{(h-1)}(l)\log [f(y_{n}|c_{n}=l;\boldsymbol{\theta})]
\\
+ p_{n}^{(h-1)}(l)\log(\pi_l) \Big].
\label{eqn:M-step1}
\end{multline}
Thus, the maximization over $\pi_l$, $l \in \cA$, can be accomplished by solving
\begin{equation}
    \begin{cases}\ds
      \max\limits_{\pi_{l}, l\in\mathcal{A}}  \sum\limits_{n=1}^{N} \sum\limits_{l \in \mathcal{A}}p_{n}^{(h-1)}(l)\log(\pi_l),\\
      \\
      \text{subject to  \quad} \sum\limits_{l \in \mathcal{A}} \pi_l = 1.
    \end{cases}
    \label{eqn:maximization1}
\end{equation}
According to the method of Lagrange multipliers, we set to zero the first derivative
with respect to the unknowns of the Langrangian function whose expression is
\begin{equation}
       \sum_{n=1}^{N} \sum_{l \in \mathcal{A}}p_{n}^{(h-1)}(l)\log(\pi_l) -\lambda \left( \sum_{l \in \mathcal{A}} \pi_l -1 \right),
    \label{eqn:lagragian1}
\end{equation}
where $\lambda$ is the Lagrange multiplier. 
Proceeding in this way, we obtain that
\begin{equation}
\begin{split}
       &\sum_{n=1}^{N} p_{n}^{(h-1)}(l) \frac{1}{\pi_l} -\lambda = 0 \\
       &\Longrightarrow \pi_l=\frac{1}{\lambda}\sum_{n=1}^{N} p_{n}^{(h-1)}(l),\quad  l\in\mathcal{A}.
\end{split}
    \label{eqn:lagragian_derivate1}
\end{equation}
Considering the constraint leads to
\begin{equation}
\begin{split}
       %&\sum_{l \in \mathcal{A}} \pi_l=
       &\frac{1}{\lambda}\sum_{l \in \mathcal{A}} \sum_{n=1}^{N} p_{n}^{(h-1)}(l) = 1\\
       & \Longrightarrow \lambda=\sum_{n=1}^{N} \sum_{l \in \mathcal{A}} p_{n}^{(h-1)}(l)=N
\end{split}
    \label{eqn:lagragian_derivate2}
\end{equation}
and, hence, 
\begin{equation}
    \pi^{(h)}_l=\frac{1}{N}\sum_{n=1}^{N} p_{n}^{(h-1)}(l), \quad l\in\mathcal{A}.
    \label{eqn:pi_l}
\end{equation}
It still remains to maximize the objective function with respect to the other parameters. Thus, neglecting
the irrelevant constants, the optimization problem to be solved can be formulated as
\begin{equation} 
\max\limits_{\boldsymbol{\theta}}  g(\boldsymbol{\theta}),
\label{eqn:maximization_f}
\end{equation}
where
\begin{multline}
g(\boldsymbol{\theta})= \sum_{n=1}^{N} \sum_{l \in \mathcal{A}}p_{n}^{(h-1)}(l)
\\
\times \left[ -\frac{1}{2}\log(2\pi\sigma_l^2) -\frac{1}{2\sigma_l^2} (y_n-a_lx_n-b_l)^2 \right].
\label{eqn:function_g}
\end{multline}
Focusing on $\sigma^2_l$, $l\in\cA$, we firstly notice that $\forall l\in\cA$
\begin{equation}
\lim_{\sigma_l^2\rightarrow+\infty} g(\boldsymbol{\theta})=-\infty
\quad \text{and} \quad \lim_{\sigma_l^2\rightarrow 0} g(\boldsymbol{\theta})=-\infty. 
\end{equation}
Thus, the maximum over $\sigma_l^2$ can be found by setting to zero 
the first derivative of $g(\boldsymbol{\theta})$ with respect to $\sigma_l^2$, namely
\begin{equation} 
 \sum_{n=1}^{N} p_{n}^{(h-1)}(l)\left[ -\frac{1}{2\sigma_l^2} +\frac{1}{2(\sigma_l^2)^2} (y_n-a_lx_n-b_l)^2 \right] =0
\label{eqn:maximization_f_sigma}
\end{equation}
and solving with respect to $\sigma_l^2$ we obtain
\begin{equation} 
\tilde{\sigma}_l^2= \frac{\sum\limits_{n=1}^{N} (y_n-a_lx_n-b_l)^2 p_{n}^{(h-1)}(l)}{\sum\limits_{n=1}^{N} p_{n}^{(h-1)}(l)},
\quad l\in\cA
\label{eqn:sigma_hat}
\end{equation}
Moreover, when $\sigma^2_l<(\tilde{\sigma}_l^2)^{(h)}$, the derivative is positive (increasing function), whereas
for $\sigma^2_l>(\tilde{\sigma}_l^2)^{(h)}$, the derivative is negative (decreasing function).
Replacing \eqref{eqn:sigma_hat} in \eqref{eqn:function_g} and neglecting the 
terms that do not enter the optimization problem, the latter is equivalent to
\begin{equation}
\min\limits_{a_l, b_l \atop l\in \mathcal{A}}  \sum_{l \in \mathcal{A}} g(a_l,b_l),
\label{eqn:minimixation_f}
\end{equation}
where
\begin{multline}
g(a_l,b_l)= \left(\sum_{n=1}^{N} p_{n}^{(h-1)}(l)\right)
\\
\times \log \left[\sum_{n=1}^{N} (y_n-a_lx_n-b_l)^2 p_{n}^{(h-1)}(l)\right].
\label{eqn:function_g_ab}
\end{multline}
Let us study the behavior of $g(a_l,b_l)$ at the endpoints of its domain. To this end, it is not difficult
to show that
\begin{equation}
\lim_{|a_l|\rightarrow +\infty \atop |b_l|\rightarrow +\infty} g(a_l,b_l)=+\infty, \quad l\in \mathcal{A}.
\end{equation}
Therefore, we set to zero the first derivative over $a_l$ of $g(a_l,b_l)$ to obtain
\begin{multline}
\left(\sum_{n=1}^{N} p_{n}^{(h-1)}(l)\right)\frac{1}{\sum\limits_{n=1}^{N} (y_n-a_lx_n-b_l)^2 p_{n}^{(h-1)}(l)}
\\
\times \left[\sum_{n=1}^{N} p_{n}^{(h-1)}(l)2(y_n-a_lx_n-b_l)(-x_n)\right]=0
\label{eqn:function_g_ab_estimation_a}
\end{multline}
and, hence,
\begin{equation}
\tilde{a}_l= \frac{\sum\limits_{n=1}^{N} p_{n}^{(h-1)}(l)(y_n-b_l)x_n}{\sum\limits_{n=1}^{N} p_{n}^{(h-1)}(l)x_n^2}, \quad l\in \mathcal{A}.
\label{eqn:function_g_ab_estimation_a1}
\end{equation}
Replacing \eqref{eqn:function_g_ab_estimation_a1} into \eqref{eqn:function_g_ab} and 
neglecting the irrelevant constants, we can consider
\be
\dmin_{b_l}\log g_1(b_l) ,
\ee
where
\begin{multline}
g_1(b_l)=\sum_{n=1}^{N} \left\{\frac{p_{n}^{(h-1)}(l)}{\left(B_l^{(h-1)}\right)^2}
\left[ y_n B_l^{(h-1)}\right.\right.
\\
-x_n\sum\limits_{\bar{n}=1}^{N} p_{\bar{n}}^{(h-1)}(l)
y_{\bar{n}}x_{\bar{n}}
\\
\left.\left.+b_l\left(x_n A_l^{(h-1)}-B_l^{(h-1)}\right)
 \right]^2\right\},
\label{eqn:function_g1}
\end{multline}
with
\begin{equation}
\begin{split}
A_l^{(h-1)}=&\sum_{\bar{n}=1}^{N} p_{\bar{n}}^{(h-1)}(l)x_{\bar{n}},
\\
B_l^{(h-1)}=&\sum_{{\bar{n}}=1}^{N} p_{_{\bar{n}}}^{(h-1)}(l)x_{\bar{n}}^2.
\end{split}
\label{eqn:B_l}
\end{equation}
Now, setting to zero the first derivative of $\log g_1(b_l)$ with respect to $b_l$ leads to
\begin{multline}
\frac{1}{g_1(b_l)}\sum_{n=1}^{N} \frac{2 p_{n}^{(h-1)}(l)}
{B^{(h-1)}_l}
\Bigg[y_n B_l^{(h-1)}
\\
-x_n\sum\limits_{\bar{n}=1}^{N} p_{\bar{n}}^{(h-1)}(l)y_{\bar{n}}
x_{\bar{n}}+b_l\left(x_n A^{(h-1)}_l-B^{(h-1)}_l\right) \Bigg] 
\\
\times \left(\frac{x_nA^{(h-1)}_l-B^{(h-1)}_l}
{B^{(h-1)}_l} \right)=0,
\label{eqn:function_g_a_estimated_estimation_b}
\end{multline}
and, as a consequence, the estimate update for $b_l$ is given by
\begin{align}
\widehat{b}^{(h)}_l &=\frac{1}{\sum\limits_{n=1}^{N} p_{n}^{(h-1)}(l)
\left(x_n A^{(h-1)}_l-B^{(h-1)}_l\right)^2}
\nonumber
\\
&\times
\Bigg\{\sum\limits_{n=1}^{N} \Bigg[ p_{n}^{(h-1)}(l)
x_n\left(x_n A^{(h-1)}_l-B^{(h-1)}_l\right)
\nonumber
\\
&\times \sum\limits_{\bar{n}=1}^{N} 
p_{\bar{n}}^{(h-1)}(l)y_{\bar{n}}x_{\bar{n}}\Bigg]-\sum\limits_{n=1}^{N} \Bigg[p_{n}^{(h-1)}(l)
y_n
\nonumber
\\
&\times\left(x_n A^{(h-1)}_l-B^{(h-1)}_l\right)\sum\limits_{{\bar{n}}=1}^{N} 
p_{_{\bar{n}}}^{(h-1)}(l)x_{\bar{n}}^2\Bigg]\Bigg\}.
\label{eqn:function_g_a_estimated_estimation_b3}
\end{align}
Finally, the updates for the estimates of $a_l$ and $\sigma^2_l$ can be written as
\begin{equation}
\widehat{a}^{(h)}_l= \frac{\sum\limits_{n=1}^{N} p_{n}^{(h-1)}(l)\left(y_n-b^{(h)}_l\right)x_n}
{\sum\limits_{n=1}^{N} p_{n}^{(h-1)}(l)x_n^2}, \quad l\in \mathcal{A}.
\label{eqn:function_g_ab_estimation_a1_1}
\end{equation}
and
\begin{equation} 
(\widehat{\sigma}_l^2)^{(h)}= \frac{\sum\limits_{n=1}^{N} \left(y_n-a^{(h)}_lx_n-b^{(h)}_l\right)^2 p_{n}^{(h-1)}(l)}
{\sum\limits_{n=1}^{N} p_{n}^{(h-1)}(l)},\quad l\in\cA,
\label{eqn:sigma_hat_1}
\end{equation}
respectively.

Summarizing, the E-step given by \eqref{eqn:E-step} and the above updates obtained from
the M-step are repeated until a stopping criterion is not satisfied. Specifically, the iterations 
end when
\begin{align}
\label{convergence_criterion}
\Delta \cL(h) =& \left| \left[\cL\left(y_{1},\dots,y_{N};\widehat{\btheta}^{(h)},\widehat{\pi}_1^{(h)},\ldots,\widehat{\pi}_L^{(h)}\right) 
\right.  \right. 
\nonumber
\\ 
&\left. \left. - \cL\left(y_{1},\dots,y_{N};\widehat{\btheta}^{(h-1)},\widehat{\pi}_l^{(h-1)},\ldots,\widehat{\pi}_L^{(h-1)}\right)\right]\right|
\nonumber
\\ 
&   / \left|\cL\left(y_{1},\dots,y_{N};\widehat{\btheta}^{(h)},\widehat{\pi}_1^{(h)},\ldots,\widehat{\pi}_L^{(h)}\right) \right| < \epsilon, 
\end{align}
where $\epsilon>0$ and
\begin{multline}
\cL\left(y_{1},\dots,y_{N};\widehat{\btheta}^{(h)},\widehat{\pi}_1^{(h)},\ldots,\widehat{\pi}_L^{(h)}\right)
\\
=\sum_{n=1}^{N} \log \left[ \sum_{l \in \mathcal{A}} f\left(y_{n}|c_{n}=l;\widehat{\btheta}^{(h)}\right)\pi^{(h)}_l\right],
\end{multline}
or after a maximum number of iterations denoted by $h_{max}$.

In the next section, we show how to incorporate the estimates obtained through the EM-algorithm into the MOS
rules to determine the number of targets $L$.

\section{Estimation Procedure for Unknown Number of Targets}
\label{sec_Unknown_L}
The estimation of the number of targets relies on the MOS rules since the hypotheses
corresponding to scenarios with different numbers of targets are nested. As stated in Section \ref{Sec:Introduction},
the maximum likelihood approach experiences an overestimation 
of the parameter space size \cite{KayLetter,Stoica1} and, hence, a penalty term is 
required to balance the corresponding growth of the likelihood function. In what follows, we consider
the Akaike Information Criterion (AIC), the Bayesian Information Criterion (BIC), and the 
Generalized Information Criterion (GIC) \cite{Stoica1}, whose general structure is
\begin{multline}
\widehat{L}=\underset{ L\in \{1,...,L_{\max}\}}{\arg\min}
\Bigg\{-2 \cL\bigg(y_{1},\dots,y_{N};\widehat{\btheta}^{(h_L)},
\\
\widehat{\pi}_1^{(h_L)},\ldots,\widehat{\pi}_L^{(h_L)}\bigg) + p(L)\Bigg\},
\label{eqn:MOS_L}
\end{multline}
where $h_L$ is the number of iterations used by the EM-based estimation procedure
introduced in Section \ref{sec_ParEst_known_L} assuming $L$ targets and
$p(L)$ is a penalty term defined as follows
\be
p(L)=
\begin{cases}
2 n_p(L), & \mbox{AIC},
\vspace{2mm}
\\
(1+\rho) n_p(L), \ \rho\geq 1, & \mbox{GIC},
\vspace{2mm}
\\
n_p(L)\log(N), & \mbox{BIC},
\end{cases}
\ee
with $n_p(L)=4L$ being the number of unknown parameters in the presence of $L$ targets.

Once $\widehat{L}$ is computed through \eqref{eqn:MOS_L}, we can define $\widehat{\cA}=\{1,\ldots,\widehat{L}\}$ and
the association rule \eqref{eqn:MAPclustering} becomes
\be
\widehat{l}=\underset{l\in\widehat{\mathcal{A}}}{\arg\max} \
P\left\{c_{n}=l|y_n;\widehat{\btheta}^{(h_{\widehat{L}})},\widehat{\pi}^{(h_{\widehat{L}})}_{l}\right\},
\label{eqn:MAPclusteringLest}
\ee
where $\widehat{\btheta}^{(h_{\widehat{L}})}$ and $\widehat{\pi}^{(h_{\widehat{L}})}_{l}$, $l\in\widehat{\cA}$, are the
estimates corresponding to the case $L=\widehat{L}$.

\section{Numerical Examples and Discussion}
\label{Sec:Analysis}
In this section, we provide illustrative examples aimed at assessing
the classification/clustering performance of the proposed approach by 
resorting to Monte Carlo (MC) counting techniques. Specifically, we start the analysis 
by assuming that $L$ is known and consider two operating scenarios with $L=5$ and $L=10$ targets. 
Then, we focus on the estimation capabilities for the number of targets by setting $L_{\max}=10$
and the actual value of $L$ equal to $3$ and $7$.

In each scenario, we directly generate the trajectories of multiple targets in 
Cartesian coordinates with the number of measurements for the $l$th target, 
$N_l$ say, $l=1,\ldots,L$,\footnote{Notice that the constraint 
$\sum\limits_{l=1}^{L} N_l = N$ holds.} being a uniformly distributed random variable in $[60, 90]$. 
In addition, trajectory intersections are included in the considered scenarios making 
the classification task more challenging.

For comparison purposes, we compare the proposed approach with two classical machine learning 
clustering techniques, namely the 
K-Nearest-Neighbor (KNN) and K-means \cite{murphy2012machine,Wang8750837,Balusamy9415318}. 
The former is a supervised method, whereas the latter is unsupervised.
In the ensuing examples, we assume that the ratio between the quantity of training and validation 
data for the KNN is $4/6$ and the number of nearest neighbors is $50$. 

As for the initialization of the EM-based procedure, we set $\pi_l = 1/L$, $l=1,\ldots,L$, while the 
initial values of $a_l$ and $b_l$ are computed by means of an {\em ad hoc} strategy summarized 
in Algorithm \ref{alg:initializationEM}. 
Possible initial values the noise variances are 
\be
\sigma_l^2= \sum\limits_{n_l^{\prime}=1}^{N_l^{\prime}} (y_{n_l^{\prime}}-a_l x_{n_l^{\prime}}-b_l)^2/N_l^{\prime},
\ l\in\cA,
\ee 
where $N_l^{\prime}$ denotes the number of measurements used to compute the initial trajectory 
coefficients of the $l$th target in Algorithm \ref{alg:initializationEM}. 

In each scenario, the measurements are affected by an uncertainty with variance 
$\sigma_l^2 = 50$, $l=1,\ldots,L$ that lead to a significant ``mix'' of the targets' measurements. 
Finally, we anticipate here that, in order to achieve a satisfactory compromise between convergence 
and computational load, $\epsilon$ is set to $10^{-5}$ and $h_{max}$ is set to $150$ and $250$ for
$L=5$ and $L=10$, respectively, as corroborated by the subsequent convergence analysis. 

\begin{algorithm}[tbp] 
	\caption{Initialization of $a_l$ and $b_l$.}
	\label{alg:initializationEM}
	\KwIn{$L$, $y_n$, $x_n$, $n = 1 ,\ldots,N$}
	\KwOut{$a_l$, $b_l$, $l = 1, \ldots, L$ }
	\textbf{Initialization:} set $N_1^{\prime} = N$, $y_{n_1^{\prime}} = y_n $, $x_{n_1^{\prime}} = x_n $, $n_1^{\prime} = n = 1 ,\ldots,N$\\
	\For{$l = 1, \ldots, L$}{
		\textbf{1.} compute $a_l$ and $b_l$ of the linear regression by $a_l = \left(\sum\limits_{n_l^{\prime} = 1}^{N_l^{\prime}}(x_{n_l^{\prime}}-\bar{x})(y_{n_l^{\prime}}-\bar{y})\right)/\sum\limits_{n_l^{\prime} = 1}^{N_l^{\prime}}(x_{n_l^{\prime}}-\bar{x})^2$ and $b_l = \bar{y} - a_l  \bar{x}$ using $N_l^{\prime}$ measurements with $\bar{x}$ and $\bar{y}$ the mean value of $x_{n_l^{\prime}}$ and $y_{n_l^{\prime}}$, respectively; \\
		\textbf{2.} compute the standard deviation of $(x_{n_l^{\prime}},y_{n_l^{\prime}})$ with respect to the fitted line, namely, $ d_l(n_l^{\prime}) = |y_{n_l^{\prime}}-a_l x_{n_l^{\prime}}-b_l|/\sqrt{a_l^2+b_l^2} $, $n_l^{\prime} = 1,\ldots,N_l^{\prime}$;\\
		\textbf{3.} set $l = l+1$;\\
		\textbf{4.} select the $\lfloor N/L \rfloor$ measurements with the minimum $d_l(n_l^{\prime})$ and discard them, we have $N_l^{\prime} = N_{l-1}^{\prime} - \lfloor N/L \rfloor$;\\
		\textbf{5.} go to step 1 using the remaining measurements until the loop ends.}
\end{algorithm}

\subsection{First Operating Scenario for Known Number of Targets $L=5$}

\begin{figure}[tbp]
	\centering
	\includegraphics[width=.45\textwidth, height=6cm]{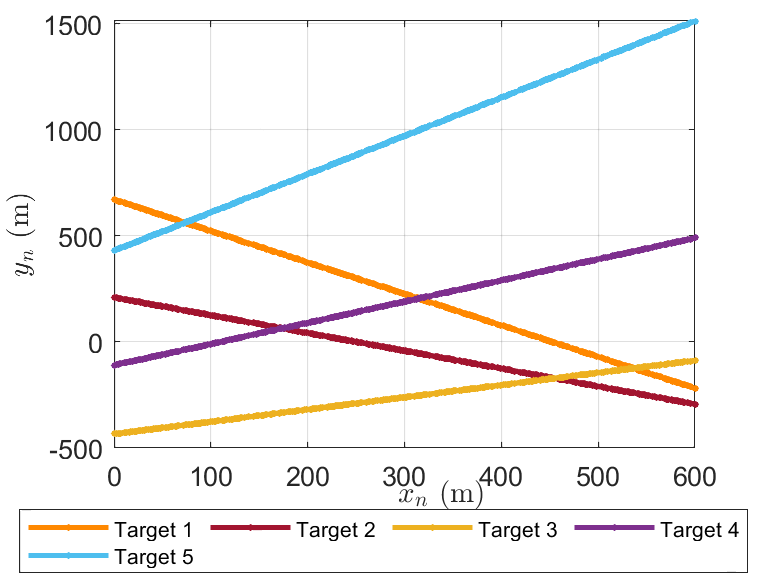}
	\caption{Clean target trajectories without measurement noise (first scenario).}
	\label{fig:trajectoriesScenario1}
\end{figure}

\begin{figure}[tbp]
	\centering
	\includegraphics[width=.45\textwidth, height=6cm]{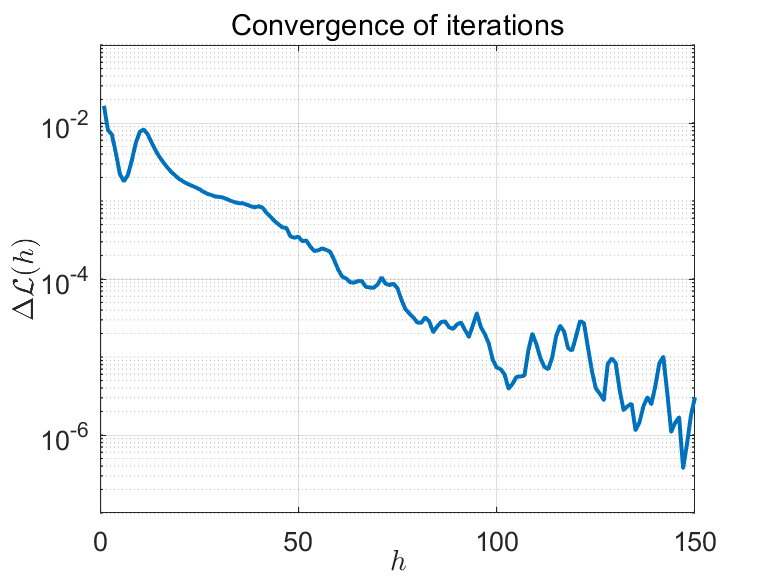}
	\caption{$\Delta \cL(h)$ versus $h$ for the EM-based procedure (first scenario).}
	\label{p1}
\end{figure}

\begin{figure*}[tb]
	\centering

		\begin{minipage}{0.4\linewidth}
			
			\centering
			\includegraphics[width=1\linewidth]{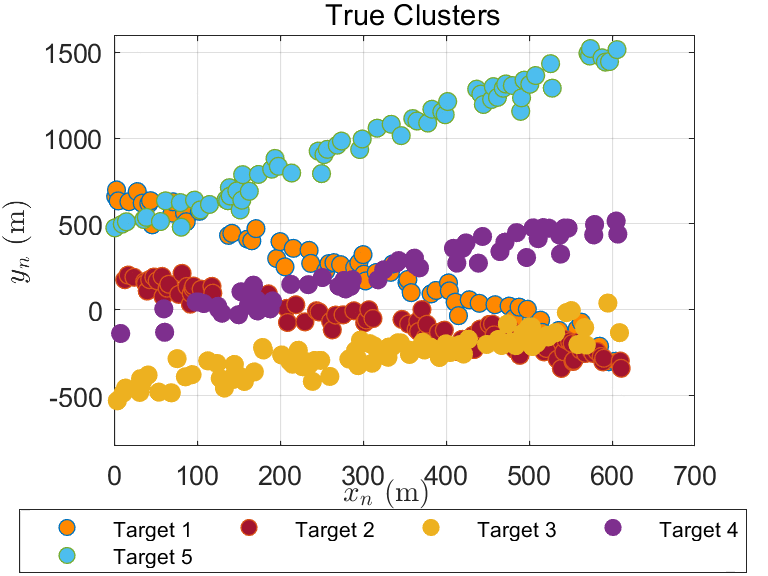}
			\caption*{(a)}
		\end{minipage}
		\begin{minipage}{0.4\linewidth}
			\centering
			\includegraphics[width=1\linewidth]{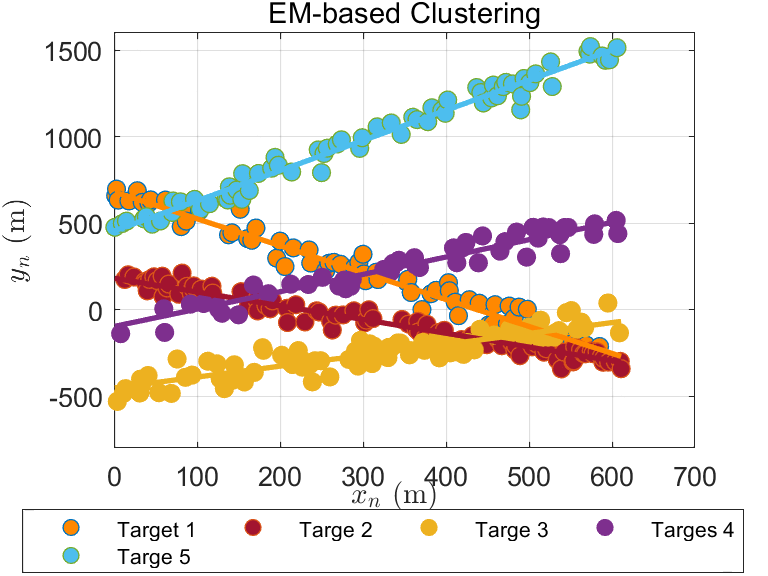}
			\caption*{(b)}
		\end{minipage}

		\begin{minipage}{0.4\linewidth}
			\centering
			\includegraphics[width=1\linewidth]{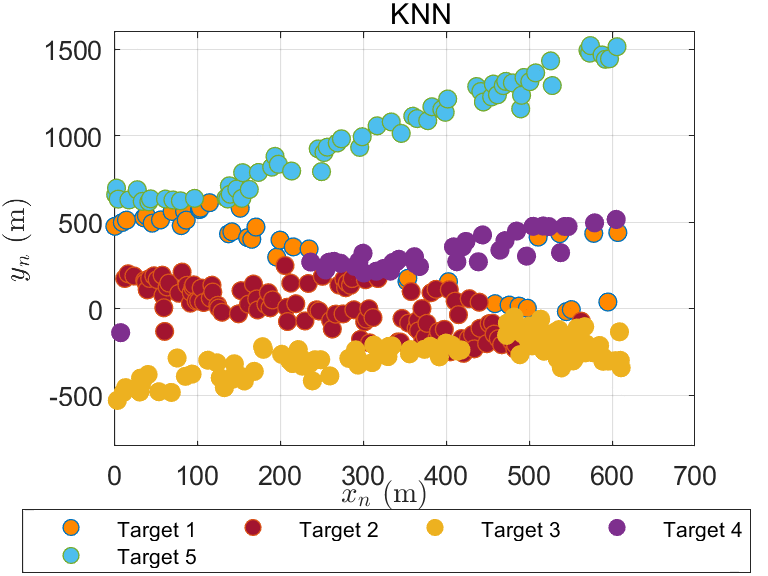}
			 \caption*{(c)}
	\end{minipage}
		\begin{minipage}{0.4\linewidth}
			\centering
			\includegraphics[width=1\linewidth]{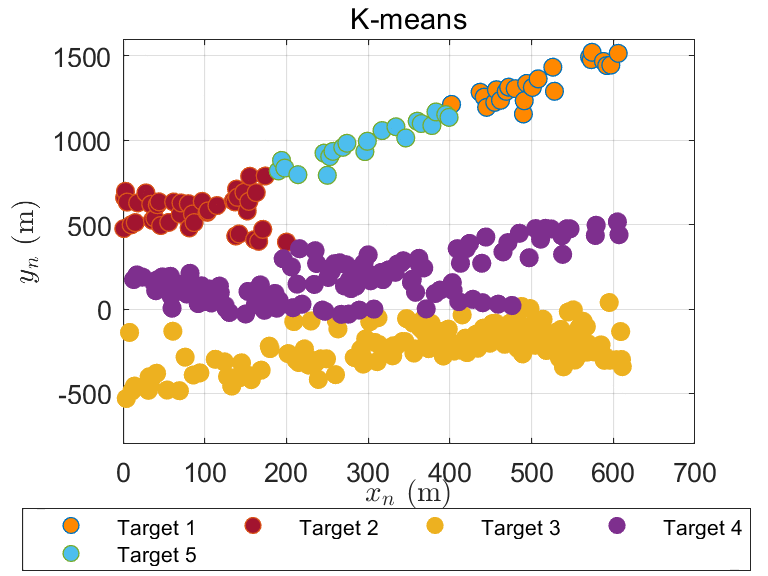}
			 \caption*{(d)}
	\end{minipage}

	\caption{Cartesian coordinates diagrams for each target over a single MC trial (first scenario):
		(a) true measurement association;
		(b) classification (scatter points) results and trajectory fitting (straight lines) for the proposed architecture;
		(c) classification results for the KNN;
		(d) classification results for the K-means.
	}
	\label{p2}
	
\end{figure*}

In this scenario, we consider five targets whose measurements are generated as follows
\begin{itemize}
	\item \textbf{Target 1}: $y_{n_1}|c_{n_1} = 1 \sim \cN(-1.4826 x_{n_1} +671,\ \sigma_1^2)$, \ $n_1 = 1,\ldots, N_1$;
	\item \textbf{Target 2}: $y_{n_2}|c_{n_2} = 2 \sim \cN(-0.8391 x_{n_2} +310,\ \sigma_2^2)$, \ $ n_2 = 1,\ldots, N_2$;
	\item \textbf{Target 3}: $y_{n_3}|c_{n_3} = 3 \sim \cN(0.5774 x_{n_3} - 434,\ \sigma_3^2)$, \ $n_3 = 1,\ldots, N_3$;
   	\item \textbf{Target 4}: $y_{n_4}|c_{n_4} = 4 \sim \cN(x_{n_4} - 110,\ \sigma_4^2)$, \ $n_4 = 1,\ldots,N_4$;
	\item \textbf{Target 5}: $y_{n_5}|c_{n_5} = 5 \sim \cN(1.8040 x_{n_5} + 430,\ \sigma_5^2)$, \ $n_5 = 1,\ldots,N_5$;
\end{itemize}
where $x_{n_l}$, $l = 1,\ldots,L$, are generated by randomly selecting integers from $1$ to $N$.
The above target trajectories without measurement noise are shown in Figure \ref{fig:trajectoriesScenario1}.

In Figure \ref{p1}, we plot the curves of $\Delta \cL(h)$ averaged over 
$1000$ independent trials to select a suitable value for $h_{max}$. It can be seen that 
the relative variation of the log-likelihood function is lower than $10^{-5}$ 
after 150 iterations and, hence, for this scenario, we set $h_{max}=150$. 

\begin{figure}[tbp]
	\centering
	\includegraphics[width=.45\textwidth, height=6cm]{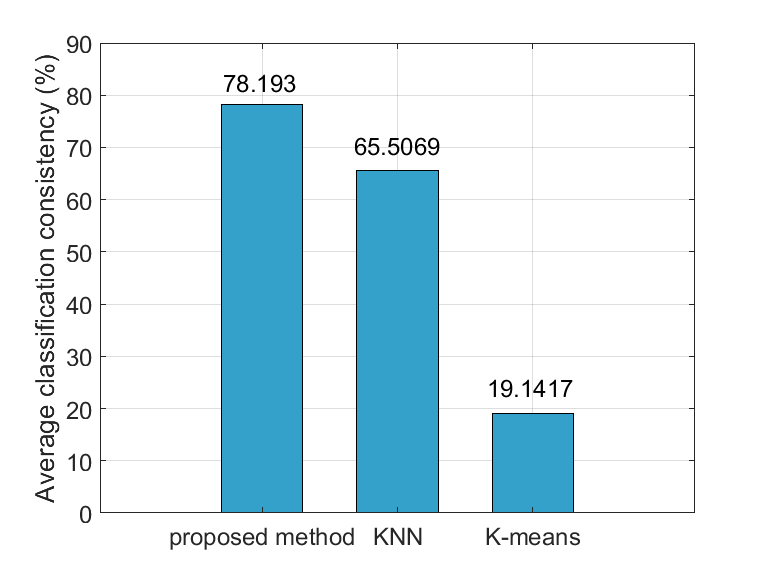}
	\caption{ Average classification consistency (\%) over $1000$ independent trials 
	for the considered classifiers (first scenario).}
	\label{p3}
\end{figure}

\begin{figure}[tbp]
	\centering
	\includegraphics[width=.45\textwidth, height=6cm]{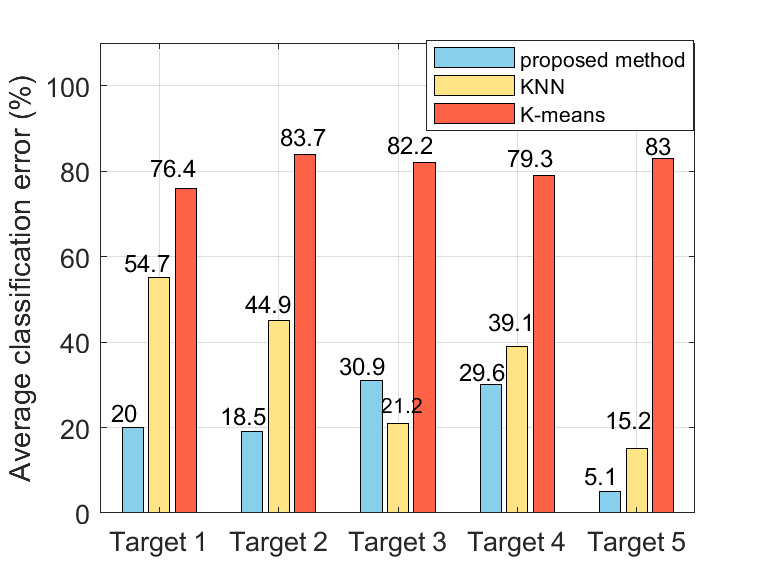}
	\caption{ Average classification error (\%) of each target  over $1000$ independent trials 
	for the considered classifiers (first scenario).}
	\label{p4}
\end{figure}

The classification capabilities of the proposed architecture in comparison with 
the two considered competitors are investigated in Figures \ref{p2}-\ref{p4}. 
Specifically, the true clusters and the classification results over a single MC trial are shown 
in Figure \ref{p2}. The inspection of this figure clearly points out 
the superiority of the proposed method in measurement labeling over the counterparts. 
As a matter of fact, from a qualitative point of view, both 
KNN and K-means experience evident misclassification errors due to the association
of measurements to wrong targets. As a consequence, the trajectory of a given target appears
divided into segments corresponding to different targets. This kind of segmentation
is more evident in Subfigure \ref{p2}(d) that reports the classification results of K-means.
Moreover, Subfigure \ref{p2}(b) contains the target trajectories, obtained through the estimates of 
$a_l$ and $b_l$, $l = 1,\ldots,L$, that\footnote{Observe that analogous curves are not reported in the other subfigures
since the KNN and K-means cannot provide such estimates.} perfectly fit with the measurements 
(at least for the considered parameter setting).

Figure \ref{p3} shows the mean classification consistency (\%), which is 
defined as the ratio between the number of correct classifications over 
the true target categories averaged over $1000$ independent trials. It turns out that 
the proposed EM-based classifier achieves a classification gain of 
approximately $12.6 \%$ and $59 \%$ with respect to the KNN and K-means, respectively. 
Figure \ref{p4} shows the mean classification error (\%) averaged over $1000$ MC trials, 
namely the ratio between the number of misclassified measurements for a given class (target) and the 
true quantity for that category. Again, the advantage of the proposed method over the considered 
competitors is quite evident. 

Finally, to assess the estimation accuracy for the estimates of $a_l$ and $b_l$ at the $n_{mc}$th MC trial, which are
generally denoted by $\widehat{a}_l(n_{mc})$ and $\widehat{b}_l(n_{mc})$, $l=1,\ldots,L$, respectively, 
in Table \ref{T1} we evaluate the Percentage Root Mean Square Error (PRMSE) relative to the true values that is defined as
\begin{equation}
\begin{cases}
\ds
\mbox{PRMSE}_{a_l} = \sqrt{\sum\limits_{n_{mc}=1}^{N_{mc}}   \frac{ \ds \min_{l^{\prime}\in\mathcal{A}} 
\left( a_l-\widehat{a}_{l^{\prime}} \right)^2}{N_{mc} } } \times \frac{100}{|a_l|},
\vspace{3mm}
\\
\ds
\mbox{PRMSE}_{b_l} = \sqrt{\sum\limits_{n_{mc}=1}^{N_{mc}} 
\frac{ \ds \min_{l^{\prime}\in\mathcal{A}} \left( b_l-\widehat{b}_{l^{\prime}}\right)^2}{N_{mc} } } \times \frac{100}{|b_l|}
\end{cases}
\label{eqn:RMSE}
\end{equation}
with $N_{mc} = 1000$ being the number of MC trials. The table highlights that
for targets $3$ and $4$ the estimate of $a_l$ gives rise to errors greater than $20 \%$ due to the fact that
the trajectories of these targets are characterized by intersections with other lines
whose angular coefficient is considerably different. The same remark also holds for what concerns
the errors related to the estimate of $b_l$. Otherwise stated, even though the percentage of correct
classification is high, when an error occurs, its value can
be high due to line intersections. Nevertheless, such errors can be mitigated by filtering
the estimates over several consecutive processed batches of measurements.

\begin{table}[tbp]
	\centering
	\caption{PRMSE values (\%) for $a_l$ and $b_l$ , $l=1,\ldots,L$, over 1000 independent trials (first scenario).}
	\begin{tabular}{|c|c|c|c|c|c| p{2.5 cm}}
		\hline
		& $l=1$ & $l=2$ & $l=3$ & $l=4$ & $l=5$ \\
		\hline
		$a_l$ &5.3967   & 17.5596  & 58.5285  & 36.8473   & 1.9468   \\           
		\hline
		$b_l$ &4.7652 &  25.8798   &  13.3593 & 45.2495  &  3.1853 \\         
		\hline
	\end{tabular}
	\label{T1}
\end{table}

\begin{figure}[tbp]
	\centering
	\includegraphics[width=.45\textwidth, height=6cm]{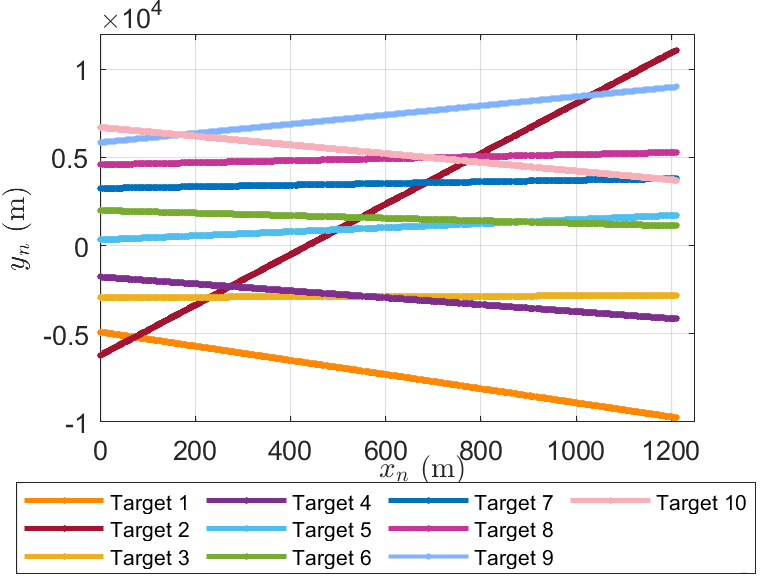}
	\caption{Clean target trajectories without measurement noise (second scenario).}
	\label{fig:trajectoriesScenario2}
\end{figure}

\begin{figure}[tbp]
	\centering
	\includegraphics[width=.45\textwidth, height=6cm]{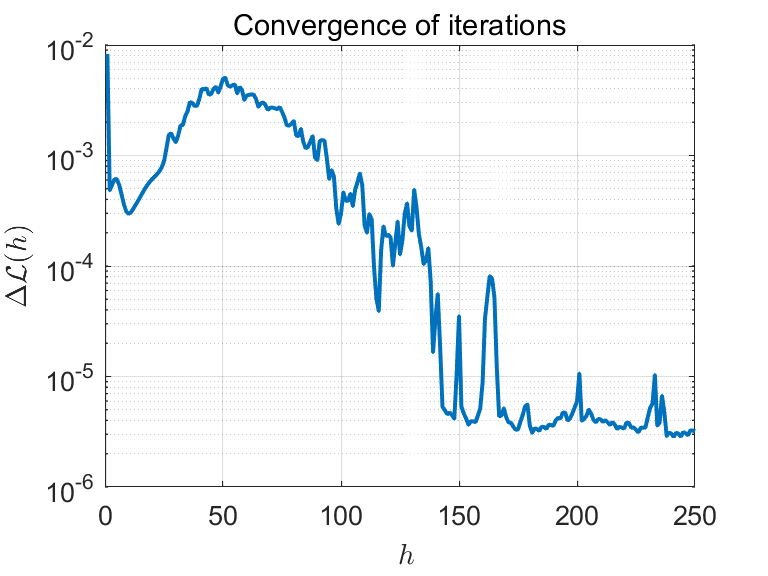}
	\caption{$\Delta \cL$ versus $h$ for the EM-based procedure (second scenario).}
	\label{p5}
\end{figure}

\begin{figure*}[tbp]
	\centering
	
	\begin{minipage}{0.4\linewidth}
		
		\centering
		\includegraphics[width=1\linewidth]{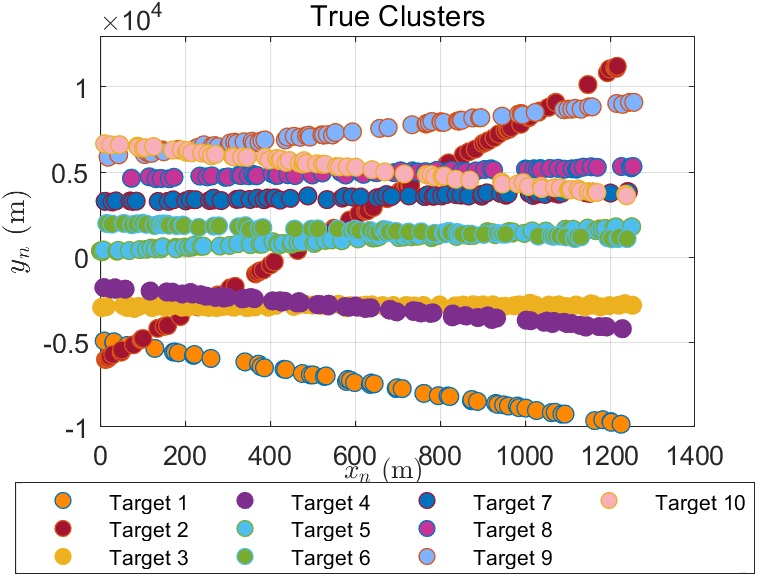}
		\caption*{(a)}
	\end{minipage}
	\begin{minipage}{0.4\linewidth}
		\centering
		\includegraphics[width=1\linewidth]{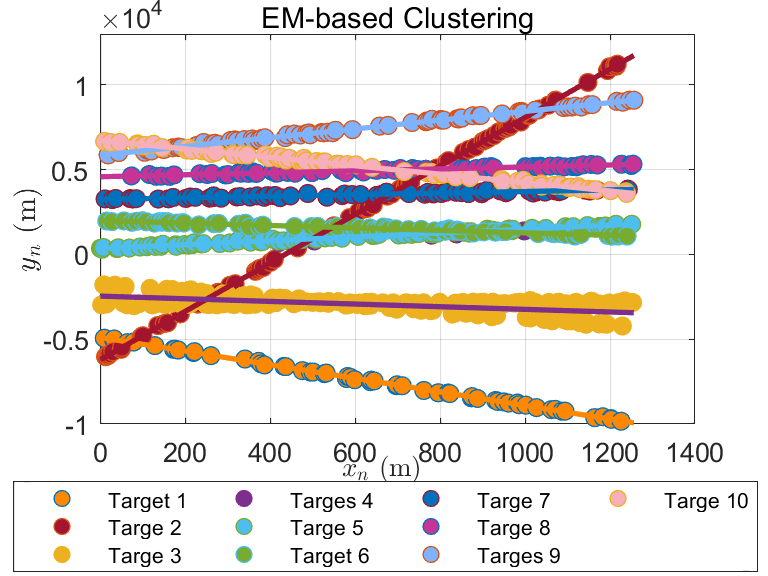}
		\caption*{(b)}
	\end{minipage}

	\begin{minipage}{0.4\linewidth}
		\centering
		\includegraphics[width=1\linewidth]{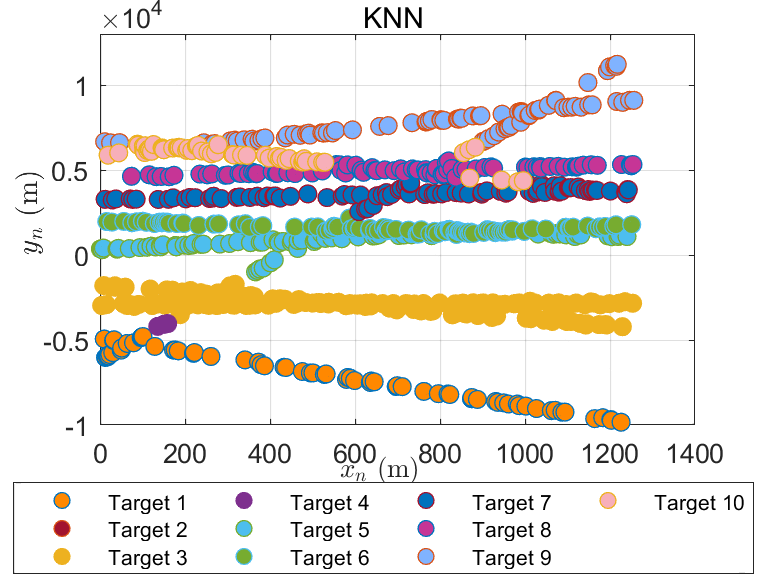}
		\caption*{(c)}
	\end{minipage}
	\begin{minipage}{0.4\linewidth}
		\centering
		\includegraphics[width=1\linewidth]{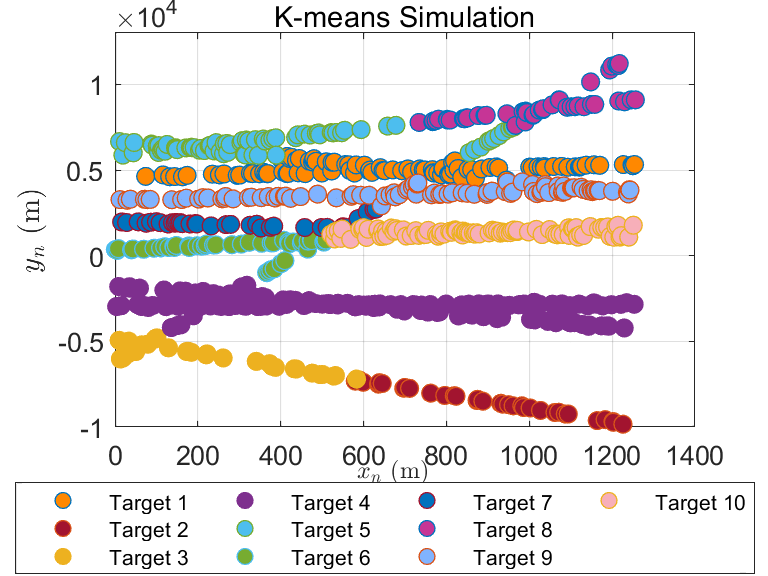}
		\caption*{(d)}
	\end{minipage}
	
	\caption{Scatter diagrams for each target over one trial (second scenario):
		(a) true measurement association;
		(b) classification (scatter points) results and trajectory fitting (straight lines) for the proposed architecture;
		(c) classification results for the KNN;
		(d) classification results for the K-means. 
	}
	\label{p6}
	
\end{figure*}

\begin{figure}[tbp]
	\centering
	\includegraphics[width=.45\textwidth, height=6cm]{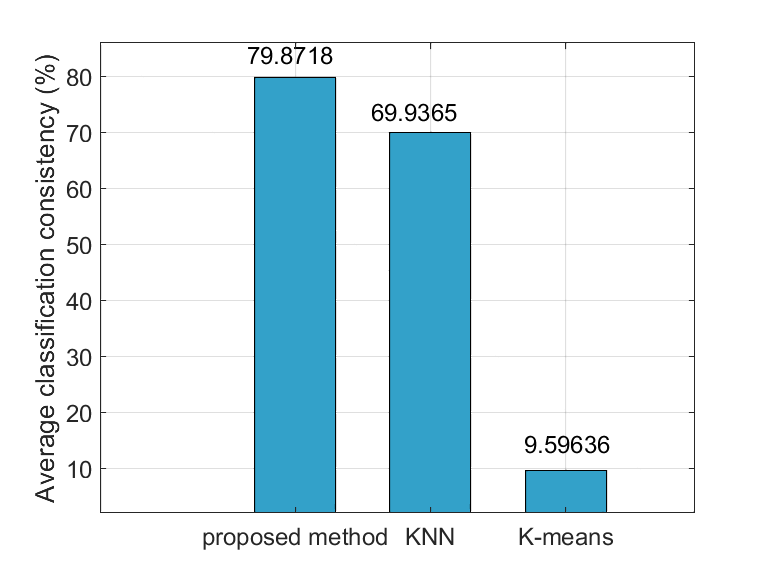}
	\caption{Average classification consistency (\%) over 1000 independent trials (second scenario).}
	\label{p7}
\end{figure}

\begin{figure}[tbp]
	\centering
	\includegraphics[width=.5\textwidth, height=6cm]{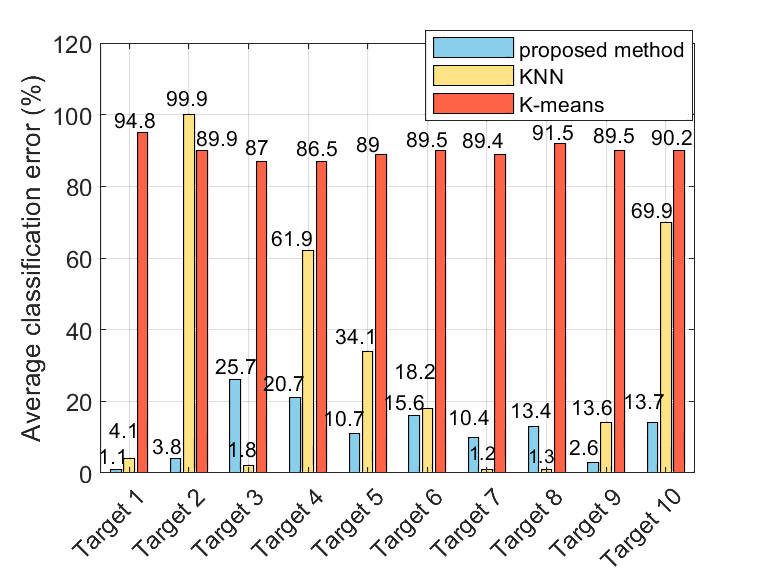}
	\caption{ Average classification error (\%) of each target  over 1000 independent trials (second scenario).}
	\label{p8}
\end{figure}

\begin{table*}[tbp]
	\centering
	\caption{PRMSE values (\%) for $a_l$ and $b_l$ , $l=1,\ldots,L$, over 1000 independent trials (second scenario).}
	\begin{tabular}{|c|c|c|c|c|c|c|c|c|c|c|c| p{2.5 cm}}
		\hline
	& $l=1$ & $l=2$ & $l=3$ & $l=4$ & $l=5$&  $l=6$ & $l=7$ & $l=8$ & $l=9$ & $l=10$\\
		\hline
		$a_l$ &5.1741  & 1.7061    & 39.4472   &  53.9560  &    23.3503  & 16.7256  &23.4816 &   21.8546  &  11.5735  &     53.5897\\ 
		\hline
		$b_l$& 0.8867 & 0.1999 &   17.5633  &  28.8967&   79.0360 &  11.5884    &     10.3934   &   10.3873   &    1.0407  &     5.7240 \\
		\hline                          
	\end{tabular}
	\label{T2}
\end{table*}

\subsection{Second Operating Scenario for Known Number of Targets $L=10$}
Now, we consider a more challenging scenario where $L=10$ targets are present with different
trajectories and intersections. As for the previous case, the KNN and K-means classifiers are taken into 
account as natural competitors. Target measurements are generated as follows
\begin{itemize}
	\item  \textbf{Target 1}: $y_{n_1}|c_{n_1} = 1 \sim \cN(-4.0108 x_{n_1} - 4897,\ \sigma_1^2)$, \ $n_1 = 1,\ldots,N_1$;
	\item  \textbf{Target 2}: $y_{n_2}|c_{n_2} = 2 \sim \cN(14.3007 x_{n_2} - 6230,\ \sigma_2^2)$, \ $ n_2 = 1,\ldots,N_2$;
	\item  \textbf{Target 3}: $y_{n_3}|c_{n_3} = 3 \sim \cN(0.0875 x_{n_3} - 2936,\ \sigma_3^2)$, \ $n_3 = 1,\ldots,N_3$;
	\item  \textbf{Target 4}: $y_{n_4}|c_{n_4} = 4 \sim \cN(-1.9626 x_{n_4} - 1774,\ \sigma_4^2)$, \ $ n_4 = 1,\ldots,N_4$;
	\item  \textbf{Target 5}: $y_{n_5}|c_{n_5} = 5 \sim \cN(1.1504 x_{n_5}+	330,\ \sigma_5^2)$, \ $n_5 = 1,\ldots,N_5$;
	\item  \textbf{Target 6}: $y_{n_6}|c_{n_6} = 6 \sim \cN(-0.7265 x_{n_6} + 1997,\ \sigma_6^2)$, \ $n_6 = 1,\ldots,N_6$;
	\item  \textbf{Target 7}: $y_{n_7}|c_{n_7} = 7 \sim \cN(0.4663 x_{n_7} + 3245,\ \sigma_7^2)$, \ $ n_7 = 1,\ldots,N_7$;
	\item  \textbf{Target 8}: $y_{n_8}|c_{n_8} = 8 \sim \cN(0.5774 x_{n_8} + 4588,\ \sigma_8^2)$, \ $n_8 = 1,\ldots,N_8$;
	\item  \textbf{Target 9}: $y_{n_9}|c_{n_9} = 9 \sim \cN(2.6051 x_{n_9} + 5846,\ \sigma_9^2)$,\ $ n_9 = 1,\ldots,N_9$;
	\item  \textbf{Target 10}: $y_{n_{10}}|c_{n_{10}} = 10 \sim \cN(-2.4751 x_{n_{10}} + 6706,\ \sigma_{10}^2)$, \ $n_{10} = 1,\ldots,N_{10}$.
\end{itemize}
The number of measurements for each target is generated as in the previous case and the clean
trajectories are shown in Figure \ref{fig:trajectoriesScenario2}.
The classification performances of the proposed architecture are investigated assuming
$h_{max} = 250$. Such a value is selected from Figure \ref{p5} where 
the relative variation of the log-likelihood is below $10^{-5}$ for $h=250$.

In Figures \ref{p6}-\ref{p8}, we provide a qualitative and quantitative assessment of 
the classification performance for the three algorithms. 
As observed for the case $L=5$, the EM-based classifier is less inclined to partition the measurement
set corresponding to a given target into subsets associated to other targets. This behavior is evident
in Figure \ref{p6} where the K-means experiences the worst performance 
corroborating what indicated by Figure \ref{p3}. 
Figures \ref{p7}-\ref{p8} point out the EM-based method provides an overall performance that is superior
with respect to that of the considered competitors, even though for some targets the KNN (that is a supervised
method) can return lower classification errors with respect ot the EM-based classifier (that is an unsupervised method).

Finally, in Table \ref{T2}, we report the PRMSE for the estimates of $a_l$ and $b_l$, $l=1,\ldots,10$. 
In this case, the errors related to $b_l$ are small except for target $5$ whose
trajectory intersects that of target $6$ maintaining a low separation. As for the errors
related to $a_l$, the highest values are returned for targets $3$, $4$, and $10$. In fact,
the lines corresponding to these targets share an intersection with the line associated to target $2$
whose angular coefficient is significantly different from the other.

\subsection{Operating Scenario where  Number of Targets is Unknown}
\begin{figure}[tbp]
	\centering
	\includegraphics[width=.45\textwidth, height=6cm]{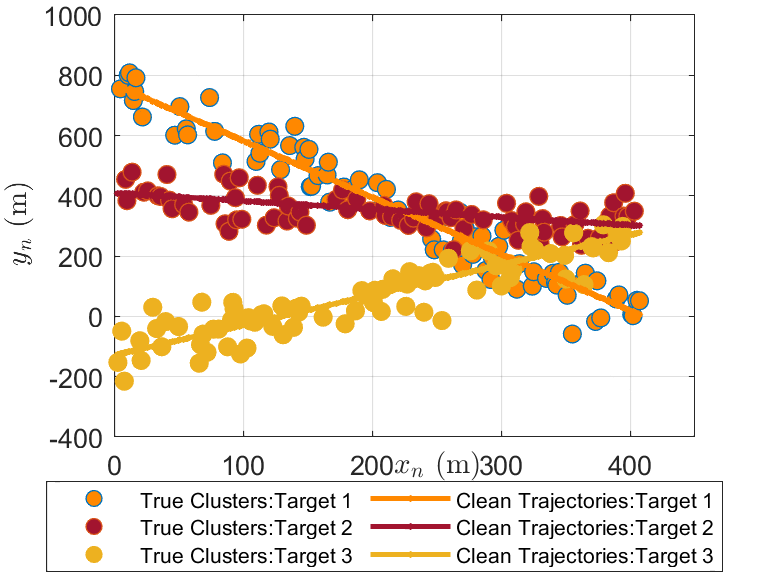}
	\caption{ True clusters and clean trajectories ($L=3$).}
	\label{p9}
\end{figure}

\begin{figure}[tbp]
	\centering
	\includegraphics[width=.45\textwidth, height=6cm]{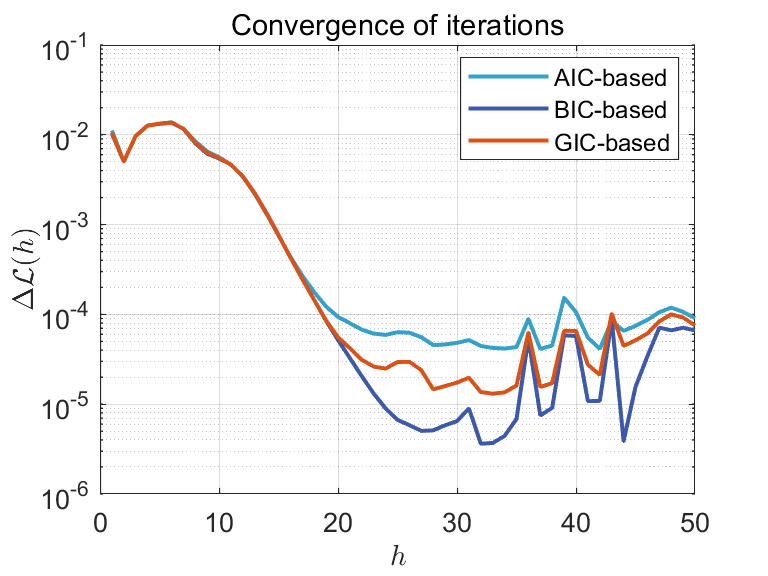}
	\caption{ $\Delta \cL$ versus $h$ of the EM procedure for the classifiers based on 
	AIC, BIC, and GIC ($L=3$).}
	\label{p10}
\end{figure}

\begin{figure}[tbp]
	\centering
	\includegraphics[width=.45\textwidth, height=6cm]{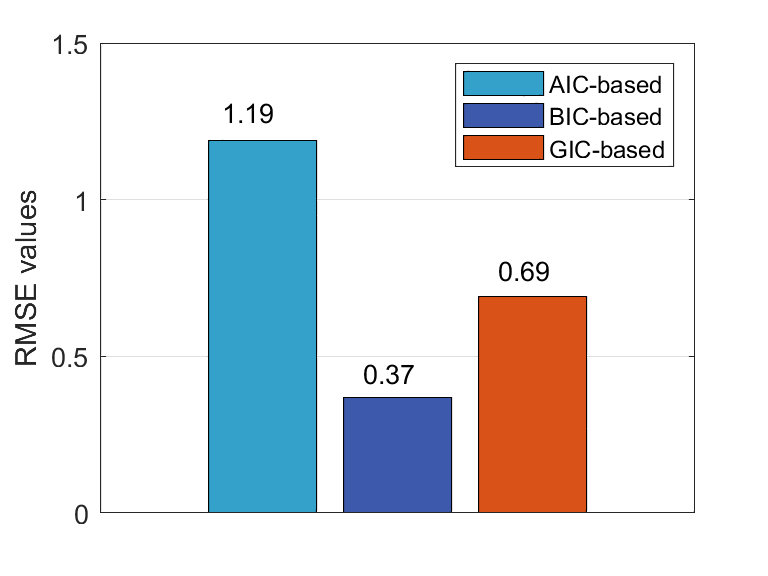}
	\caption{RMSE values for $L$ for the classifiers based upon the AIC, BIC, 
	and GIC over 1000 independent trials ($L=3$).}
	\label{p11}
\end{figure}

In this section, the classification and estimation performance is assessed 
when the number of targets is unknown under the constraint $L_{\max}=10$. 
The scenario considered in what follows comprises $L=3$ targets whose trajectories are
\begin{itemize}
\item \textbf{Target 1}: $y_{n_1}|c_{n_1} = 1 \sim \cN(-1.8807 x_{n_1} + 771,\ \sigma_1^2)$, \ $n_1 = 1,\ldots, N_1$;
\item \textbf{Target 2}: $y_{n_2}|c_{n_2} = 2 \sim \cN(-0.2679 x_{n_2} +410,\ \sigma_2^2)$, \ $ n_2 = 1,\ldots, N_2$;
\item \textbf{Target 3}: $y_{n_3}|c_{n_3} = 3 \sim \cN(x_{n_3} - 129,\ \sigma_3^2)$, \ $n_3 = 1,\ldots, N_3$,
\end{itemize}
More specifically, true clusters with measurement noise variance $\sigma_l^2=50$ and 
the clean trajectories are shown in Figure \ref{p9}. 
Before presenting the estimation results, the convergence rate of 
the EM procedure over $1000$ independent trials using AIC, BIC, and GIC is depicted in Figure \ref{p10}. 
The results confirm that $h_{max}=50$ returns a relative variation of $\Delta\cL$ lower
than $10^{-4}$. The estimation performance related to $a_l$, $b_l$, and $L$ is investigated using 
the metrics defined in \eqref{eqn:RMSE} and 
\begin{equation}
\mbox{RMSE}_{L} = \sqrt{\sum\limits_{n_{mc}=1}^{N_{mc}}    \ds  
\left( \widehat{L}(n_{mc})-L \right)^2/N_{mc}  }
\label{rmse}
\end{equation}
with $N_{mc}=1000$. 
Table \ref{T3} contains the PRMSE values for what concerns the estimation of the trajectory parameters 
and points out that the BIC-based clustering architectures
provide better results than the classifiers based upon the AIC and GIC.
In Figure \ref{p11}, the RMSE values associated with the estimation of $L$ 
confirm the superiority of the BIC-based classifier that returns an error lower than $0.5$ 
at least for the considered parameters.

\begin{table}[tbp]
	\centering
	\caption{PRMSE values (\%) for $a_l$ and $b_l$ , $l=1,\ldots,L$, using AIC, BIC, and GIC over 
	1000 independent trials ($L=3$).}
	\renewcommand\arraystretch{1.35}
	\resizebox{\linewidth}{!}{
	\begin{tabular}{|c c|c|c|c|}
		\hline
		&  &$l=1$ & $l=2$ & $l=3$  \\
		\hline
		\multirow{2}{*}{\textbf{AIC-based} } & $a_l$ &4.1803  & 30.1187  &  8.6034\\       
		                                     & $b_l$ &1.8741  &  3.7893  & 11.9821 \\        
		\hline
		\multirow{2}{*}{\textbf{BIC-based} } & $a_l$ &3.8201  & 27.1248 &  7.5070\\       
		                                     & $b_l$ & 1.7480   & 3.2314  & 10.1665 \\         
		\hline
		\multirow{2}{*}{\textbf{GIC-based} } & $a_l$ & 3.8727  & 27.4196  &  7.7718\\       
	                                         & $b_l$ &   1.7651  &  3.2885  & 10.4805 \\ 
		\hline
	\end{tabular} }
	\label{T3}
\end{table}

\section{Conclusions and Future Works}
\label{Sec:conclusions}
In this work, we have proposed a solution for clustering data generated by the nodes of a radar network
where each node has limited processing capabilities. In fact, we have assumed that the fusion center
collects ($2$-dimensional) position measurements without any side information that can be used to create clusters associated
with the targets in the region of interest. To this end, we have used fictitious nonobservable random variables
that represent the label of each measurement. Then, we have estimated the posterior probability that a label
takes on a specific value given the corresponding measurement by resorting to the EM-algorithm. The clustering is performed
by selecting the label that returns the highest posterior probability. This method is clearly less time demanding
than the plain maximum likelihood approach whose computational load depends on the total number of data partitions
as well as the number of targets. The performance assessment has been conducted by using synthetic data and
in comparison with well-known data-driven clustering algorithms such as the KNN and K-means.
Different challenging scenarios have been considered and for each of them, the proposed algorithm
is capable of outperforming the considered competitors for what concerns the clustering and estimation capabilities.

Future research tracks might encompass the extension of the proposed approach to the case where
the measurements contain information related to a third-dimension or the target Doppler frequency.
The validation of the proposed approach with real recorded data from a radar network is part of the current research activity.

\bibliographystyle{IEEEtran}
\bibliography{group_bib}

\end{document}